\documentstyle[pre,aps,epsf,float,multicol]{revtex}
\def\k{\kappa}
\def\l{\lambda}
\def\d{\delta}
\def\ds{\displaystyle}
\def\tsty{\textstyle}
\def\pr{\prime}
\def\br#1{\langle #1\rangle}
\def\b#1{{\bf #1}}
\def\D{\Delta}

\def\cF{{\cal F}}
\def\cG{{\cal G}}
\def\vd#1#2{\b #1\kern -0.5mm\cdot\kern -0.5mm\b #2}
\def\mns#1#2{#1\kern -0.5mm - \kern -0.5mm #2}
\def\pls#1#2{#1\kern -0.5mm + \kern -0.5mm #2}
\begin{document}
\draft
\title{\bf Density Fluctuations in an Electrolyte from Generalized
Debye-H\"uckel Theory}
\author{Benjamin P. Lee and Michael E. Fisher}
\address{Institute for Physical Science and Technology, The University of
Maryland, College Park, Maryland 20742}
\date{\today}
\maketitle
\begin{abstract}
Near-critical thermodynamics in the hard-sphere (1,1) electrolyte is well
described, at a classical level, by Debye-H\"uckel (DH) theory with ($+$, $-$)
ion pairing and dipolar-pair-ionic-fluid coupling.  But DH-based theories
do not address {\it density\/} fluctuations.  Here density correlations are
obtained by functional differentiation of DH theory {\it generalized\/}
to {\it non}-uniform densities of various species.  The correlation length
$\xi$ diverges universally at low density $\rho$ as $(T\rho)^{-1/4}$
(correcting GMSA theory).  When $\rho=\rho_c$ one has $\xi\approx\xi_0^+/
t^{1/2}$ as $t\equiv(T-T_c)/T_c\to 0+$ where the amplitudes $\xi_0^+$
compare informatively with experimental data.
\end{abstract}
\begin{multicols}{2}
There is a major puzzle in the theory of fluid {\it criticality\/} in
model ionic systems \cite{revs} because experiments \cite{revs,expmt}
reveal that certain electrolyte solutions exhibit {\it classical}
(i.e., van der Waals as against the usual Ising-type) critical
exponents over as much as 1 to 3 decades when $|t|\equiv|T-T_c|/T_c\to
0$.  Probably there is always a scale $t_\times$ {\it below} which the behavior
crosses over from classical to Ising; but attempts to explain how $t_\times$
might vary from $\sim 1$ to $\sim 10^{-4}$ have so far been unconvincing.
Initial efforts have addressed the
simplest case: the `\b Restricted \b Primitive \b Model,' consisting
of $N=N^++N^-\equiv V\rho$ hard-sphere ions of diameter $a$, carrying
charges $\pm q$ in a medium of dielectric constant $D$.  The hope has
been to decide the universality class (and crossover scale $t_\times$
if appropriate) of the RPM \cite{revs,FL}.

To that end Fisher and Levin \cite{FL} have shown that the original
Debye-H\"uckel (DH) theory \cite{DH} provides a remarkably good, albeit
classical account of the critical thermodynamics as judged by current
simulations \cite{revs,FL}.  However, for a satisfactory description, 
pure DH theory must be extended (i), following Bjerrum \cite{revs,FL},
by recognizing bound,
neutral but dipolar (+,--) {\it ion pairs\/} in equilibrium with the free
ions, (ii) by including the dipolar-ionic (DI) solvation free energy
\cite{revs,FL}, and (iii) by allowing for hard-core (HC) repulsions.  In
terms of $T^*\equiv k_BTDa/q^2$, these DHBjDIHC theories yield
critical points in the range $T_c^*\simeq 0.052$ to
$0.057$ as compared with $0.052-0.056$ from recent simulations [3(b)].

Now, following Ebeling and Grigo \cite{EG}, one can also pursue theories
based on the mean spherical approximation (MSA); but, for reasons
currently obscure, such theories, even when improved in various ways
[3(b),5-8] yield estimates for $T_c^*$ too high by 35-50\% \cite{gmsa}.
Note also that the hypernetted chain (HNC) and other integral equations have no
solutions in the critical region!
Further study of the DH-based theories is thus well justified.

To understand properly the nature of a critical point one must go
beyond thermodynamics to study the order-parameter fluctuations.  But,
even for ionic fluids, the order parameter is just
the overall density.  Now DH-based theories illuminate the
Debye-screening of the bare Coulombic potential, as seen in the
exponential decay of the {\it charge-charge\/} correlations, but,
unfortunately, they say little about the overall {\it density-density
correlation function,\/} $G_{\rho\rho}(\b r)\equiv\br{\rho(\b r)\rho(\b
0)}-\rho^2
\equiv \rho\bigl\{\d(\b r)+\rho[g_2(\b r)-1]\bigr\}$.
Our aim here is to rectify this deficiency.

Note, especially, that the Fourier transform of $G_{\rho\rho}(\b r)$
yields the \b k-dependent susceptibility
\begin{equation}
\chi(\b k)=\hat G_{\rho\rho}(\b k)/\rho=\chi(0)/[1+\xi^2k^2+\ldots]
\end{equation}
which diverges at $\b k=0$ at criticality.  Indeed, $\chi(\b k)$
determines the critical opalescence and turbidity \cite{expmt} and 
specifies the (second-moment) correlation length $\xi(T,\rho)$ which
diverges as $\xi_0^+/t^\nu$ when $t\to 0+$ at $\rho=\rho_c$;
furthermore, $\chi(\b k)$ approaches the reduced
compressibility $\chi(0)=\rho k_BTK_T$ (or its solution analog) when
$k\to 0$ \cite{MM}.

As stressed by Fisher and Levin [3(a),7], it is valuable to
know the critical amplitude $\xi_0^+$ even {\it within} a classical 
approximation
since, via the Ginzburg criterion, that offers a possible route for 
estimating a crossover range, $\pm t_\times$, outside which close-to-classical
critical behavior might be realized \cite{GC}.

In this Letter we show how DH theory can be generalized so as to yield,
in a natural way, density fluctuations diverging at criticality
\cite{FLL,GE,Stell}.  The method extends straightforwardly to the full
DHBjDIHC theories \cite{FL} as we illustrate below.  In particular,
we calculate the correlation length $\xi(T,\rho)$ {\it explicitly\/} within the
simplest generalized (GDH) theory, and numerically, at improved levels of
approximation.  At {\it low\/} densities a novel, {\it universal divergence}
of $\xi(T,\rho)$ is predicted for all $T$.  In the
critical region the results are, as expected, classical with
$\nu={1\over 2}$; but the amplitudes $\xi_0^+$ prove informative and
are compared with experimental data \cite{expmt} in classical and
Ising domains \cite{GC}. 

Explicitly we proceed, following \cite{FL}, by approximating the total 
Helmholtz free energy $F(T,\rho)$ by a sum of terms representing ideal gas,
ionic fluid, dipole-ion, and hard-core contributions; but we now aim
for a {\it functional} $\beta F[\{\rho_j\}]=\int d^dr{\cal F}$ where
 $\rho_1(\b r)=\rho_+(\b r)+\rho_-(\b r)$ and
$\rho_2(\b r)$ are slowly varying local free-ion and dipolar pair 
densities, while
$\beta=1/k_BT$. Since we wish to probe only the density fluctuations,
we follow DH theory and maintain electroneutrality, $\rho_+(\b r)=
\rho_-(\b r)$, on long length scales.
Of central concern are the (reduced) direct correlation functions
which follow by functional differentiation (with $i,j=1,2$) as
\begin{equation}
C_{ij}(\b r-\b r^\pr)\equiv\d^2\beta F\big/\d\rho_i(\b r)\d\rho_j(\b r^\pr)
\big|_{\rho_\l(\b r^{\pr\pr})=\bar\rho_\l},
\end{equation}
where the $\bar\rho_\l$ ($\l=+,-,2$) are the {\it overall equilibrium 
densities}.  Note
that the various terms in $\cal F$ contribute {\it linearly} to the $C_{ij}$
and, in particular, $\hat C_{ij}^{\rm Ideal}(\b k)=\d_{ij}/\bar\rho_j$.
However, since the {\it total} local density is $\rho(\b r)=\rho_1(\b r)
+2\rho_2(\b r)$ \cite{dipcom} one finds, with the aid of the 
Ornstein-Zernike (OZ) matrix relation for $\hat C_{ij}(\b k)$ \cite{MM}, 
the result
\begin{equation}
{1\over\rho\chi(\b k)}={1\over \hat G_{\rho\rho}(\b k)}=
{\hat C_{11}(\b k)\hat C_{22}(\b k)-[\hat C_{12}(\b k)]^2\over 
4\hat C_{11}(\b k)-4\hat C_{12}(\b k)+\hat C_{22}(\b k)},
\end{equation}
from which $\xi$ follows by expansion in $\b k$.  More expeditiously one
may impose infinitesimal density variations
\begin{equation}\label{denvar}
\rho_j(\b R)=\bar\rho_j[1+\D_j\cos\vd k r],
\end{equation}
and expand the reduced free-energy density $\beta F/V$ in powers of
the $\D_j$: the quadratic term is then ${1\over 4}\sum_{i,j}\bar\rho_i
\bar\rho_j\hat C_{ij}(\b k)\D_i\D_j$, from which the $\hat C_{ij}$ follow.

Evidently, the crucial issue is to extend DH theory to nonuniform but 
slowly varying 
mean densities of the various species.  Note first that the free-ion 
contribution becomes \cite{long}, via the Debye charging process \cite{DH},
\begin{equation}\label{fdh}
F^{DH}=\int d^dr_1\rho_1(\b r_1)\int_0^q dq_1\psi_1(\b r_1,q_1),
\end{equation}
where $\psi_1(\b r_1;q_1)$ is the mean electrostatic potential at the site
$\b r_1$ of a fixed ion due to all the other ions when each carries
charges $\pm q$.  If $\phi(\b r,\b r_1)$ is the mean electrostatic
potential at a general point $\b r$ when the ion $1$ is fixed at $\b r_1$,
one has \cite{DH} $\psi_1(\b r_1)=\lim_{\b r\to\b r_1}[\phi(\b r;
\b r_1)-q_1/D|\b r-\b r_1|]$.  DH derived their celebrated equation 
for $\phi$ by approximating the probability
density for a particle of species $\l$ ($=+,-,2$) with charge $q_\l$ by
$\bar\rho_\l\exp[-\beta q_\l\phi(\b r)]\simeq\bar\rho_\l[1-\beta q_\l
\phi(\b r)]$ \cite{DH}.  (Note $q_2=0$.)
In the same spirit we now propose to replace the constant
partial (species) density $\bar\rho_\l$ by $\rho_\l(\b r)$ the (slowly
varying) {\it non}-uniform density \cite{long}.  Our generalized (GDH)
equation then reads
\begin{equation}\label{nudh}
[\nabla^2_r-\tilde\k^2(\b r)\Theta_1(\b r)]\phi(\b r;\b r_1)=
-4\pi q_1\d(\b r-\b r_1)/D,
\end{equation}
where, utilizing $\theta(y)$, the Heaviside step function,
$\Theta_1(\b r)\equiv\theta(|\b r-\b r_1|-a)$ embodies the 
crucial hard-core boundary condition \cite{DH}, while the spatially
varying coefficient
\begin{equation}\label{kappatil}
\tilde\kappa^2[\{\rho_j\}]=4\pi\beta\tsty\sum_\l q_\l^2\rho_\l(\b r)/D
=4\pi\beta q^2\rho_1(\b r)/D,
\end{equation}
reduces to the standard expression for $\k^2$, the inverse Debye length
squared, when $\rho_\l(\b r)=\bar\rho_\l$ is constant \cite{DH}.

To solve (\ref{nudh}), we adopt (\ref{denvar}) and expand $\phi$ in powers 
of $\D_1$.
The coefficient $\phi_n(\b r,\b r_1)$ of $\D_1^n$ can be computed 
recursively, setting $\tilde\k=\k$ and  $\b r_1=0$, with the aid of the 
Green's function
\begin{equation}\label{gdef}
\cG(\b r;\b r^\pr)={\k\over 4\pi}\sum_{\ell=0}^\infty\cG_{\ell}(\k r,
\k r^\pr)P_\ell\biggl({\vd r r^\pr\over rr^\pr}\biggr),
\end{equation}
where, employing modified spherical Bessel functions, 
\begin{equation}
\hbox{\vbox{
\halign{\hfil $#$&$#$&$#$\hfil &$\quad#$\hfil\cr
\cG_\ell(s,s^\pr)&=&\ds{s_<^\ell\over s_>^{\pls\ell 1}}-\ds{k_{\mns\ell 1}(x)
s^\ell s^{\pr\ell}\over k_{\pls\ell 1}(x)x^{2\pls\ell 1}},&s,s^\pr<x,\cr 
\noalign{\medskip}
&=&\ds(2\pls\ell 1){s_<^\ell k_\ell(s_>)\over k_{\pls\ell 1}(x) 
x^{\pls\ell 2}},& s_<\le x\le s_>,\cr}}}
\end{equation}
where $x=\k a$, $s_>=\max(s,s^\pr)$, $s_<=\min(s,s^\pr)$, while,
\begin{equation}
{\cG_\ell(s,s^\pr)\over 2\pls\ell 1}={i_{\pls\ell 1}(x)\over 
k_{\pls\ell 1}(x)}k_\ell(s)k_\ell(s^\pr)+i_\ell(s_<)k_\ell(s_>),
\end{equation}
for $s,s^\pr>x$, and $P_\ell(\mu)$ denotes a Legendre polynomial.

Substituting in (\ref{fdh}) and expanding 
$\hat C_{\rho\rho}\equiv 1/\hat G_{\rho\rho}$ to $O(k^2)$
yields $\xi^2$.  This requires only the $\ell=0$ and $1$ 
terms in (\ref{gdef}).  Consequently, within pure DH theory, the
simplest level of approximation, we find the correlation length is
given explicitly by (recalling $x=\k a$)
\begin{equation}\label{xidh}
{\xi^2\over a^2}={\chi^{DH}(0)\over 24T^*x^2}\biggl[\ln{(\pls 1 x)^{10}\over
(\pls{\pls 1 x}{{1\over 3}x^2})^9}-{\mns{\mns x 5x^2}8x^3\over 2(1+x)^2}
\biggr],
\end{equation}
where $1/\chi^{DH}(0)=\mns 1 x/4T^*(\pls 1 x)^2$ \cite{FL}.  

Now corrections to this result enter only in $O(x^2)=O(\rho)$,
i.e., beyond the leading low-density behavior which, in fact, exhibits 
the novel {\it divergence\/}
\begin{equation}\label{xilowden}
\xi(T,\rho)=\tsty{1\over 4}(b/36\pi\rho)^{1/4}[1+{1\over 8}\k b+O(\rho^*)],
\end{equation}
when $\rho\to 0$, where $b=q^2/Dk_BT$ is Bjerrum's length.  This
expression for the density-density correlation length is
{\it independent\/} of the hard-core diameter $a$ and is thus 
{\it universal}\,!  We believe it represents the {\it exact\/}
limiting behavior not previously noted.
At low densities Debye's screening length $\xi_D=1/\k$ 
controls the decay of the charge correlations [4(b),17].  It
also diverges universally when $\rho\to 0$;
but since we find $\xi\approx\sqrt{b\,\xi_D/48}$, the density
correlations then decay on a {\it shorter\/} scale than the charge 
correlations.

Our conclusion (\ref{xilowden}) can be checked further by using the
HNC relation $c_{ij}\approx-\beta u_{ij}+{1\over 2}h_{ij}^2$
[4(b),17,18], which is probably generally valid 
in the low density limit \cite{KM}
when $h_{ij}\equiv g_{ij}-1$
decays exponentially fast, as expected here.  This leads to \cite{long,Ennis}
\begin{equation}\label{chihnc}
1/\chi(\b k)\approx 1-\tsty {1\over 2}\k^2 b\tan^{-1}(k/2\k)/k\qquad(\rho
\to 0),
\end{equation}
which, on expansion to order $k^2$, yields the DH limiting law for
$\chi(0)$ and reproduces (\ref{xilowden})! 
However, the {\it true correlation length}, $\xi_\infty(T,\rho)$,
that determines the OZ-like exponential decay of $G_{\rho\rho}(\b r)$ is
determined by the dominant zeros of (\ref{chihnc}).  These give
the different expression
\begin{equation}\label{xiinf}
\xi_\infty\approx\tsty{1\over 2}\xi_D\bigl\{1+2\exp[-4/(\pi\rho b^3)^{1/2}]
+$\ldots$\bigr\},
\end{equation}
[16,19] which diverges as $(T/\rho)^{1/2}$. Thus $G_{\rho\rho}
(\b r)$ has
a small but longer-range tail decaying slightly {\it more slowly\/}
than $e^{-2\k r}/r^2$, the {\it squared\/} charge-charge correlation.

By contrast to (\ref{xilowden}) and (\ref{chihnc}), one finds
\cite{long} that the GMSA or generalized mean-spherical
approximation \cite{Evans,HS} predicts
$1/\chi(\b k)\approx 1-\tsty{1\over 2}\k b/[2+k^2 a/\k]$
when $\rho\to 0$.  This gives $\chi(0)$ correctly to $O(\rho^{1/2})$
but leads to 
\begin{equation}
\xi_{GMSA}\approx\xi_{\infty GMSA}\approx(\tsty{1\over 2} a\xi_D)^{1/2}
=\tsty{1\over 2}(a^2/\pi\rho b)^{1/4}.
\end{equation}
Thus $\xi_{GMSA}$ also diverges as $\rho^{-1/4}$, but the power of $T$ differs
and the result is {\it non}-universal, depending on $a$. This reveals an
unsuspected defect of the GMSA \cite{Enniscom}, which was especially devised to
satisfy a variety of correlation-function sum rules \cite{HS}.  (The
original MSA  gives only a hard-sphere result for $G_{\rho\rho}(\b r)$:
see, e.g.,\ \cite{MM}.)

In the critical region the pure GDH result (\ref{xidh}) diverges with 
exponent $2\nu=1$ at $T_c^*={1\over 16}$, $x_c=1$, $\rho^*_c=1/64\pi$.
The correlation length amplitude is found to be
\begin{equation}
(\xi_0^+/a)_{DH}=\bigl[\tsty 1+{20\over 3}\ln 2 - 6 \ln{7\over 3}\bigr]^{1/2}
\simeq 0.7329.
\end{equation}
This is surprisingly close to the GMSA value $0.75$ \cite{Evans}, although 
$T_c^*$ and $\rho_c^*$ differ significantly \cite{gmsa}.

However, although the pure GDH theory based on
(\ref{fdh})-(\ref{kappatil}) is sufficient at low density, one must,
as mentioned in the introduction \cite{FL}, include {\it ion pairing\/}
to study the critical region.  Bjerrum's ansatz for the association
constant is appealing but Ebeling's result is superior \cite{FL} and used
here.  (Near criticality the numerical changes are minor.)  In simple
``DHBj'' theory the ion pairs are supposed {\it ideal\/} \cite{FL} and
one finds $\hat C_{22}=1/\bar\rho_2$, $\hat C_{12}=0$, and $\hat C_{11}(\b k)$
is unaltered.  But that is too naive and proves unphysical:
it is essential to include the {\it dipole-ionic fluid\/} (DI)
{\it interactions} \cite{FL}.

We calculate the new {\it non-}uniform DI effects by using the GDH equation, 
(\ref{nudh}), but with a dipolar source term, i.e., $+$ and $-$ point 
charges at $\b r_1=\pm{1\over 2}\b a_1$, where
$\b a_1$ specifies the orientation and typical charge separation, $a_1(T)
\equiv|\b a_1|$ \cite{FL}.  The associated bispherical exclusion
zone is approximated by a sphere of radius $a_2$ \cite{FL}.  Thus
the Green's function (\ref{gdef}) still applies, but with $x\to x_2
\equiv\k a_2$.  At low $T$, $a_1=a$ (``contact'') and $a_2=1.16198a$
(angular average) are reasonable \cite{FL} and the
sensitivity to these values is readily tested.

Angular integration over the dipole orientations is complicated, yielding 
the solution $\phi_{dip}(\b r;\b r_1)$ as a multiple sum with
Clebsch-Gordan coefficients, ${\cal C}_{\ell_1,\ell_2}(m_1,m_2|\ell, m)$.
To obtain $\psi_2(\b r;q)$ for use in the pair analog of (\ref{fdh}),
the self-potential of the source dipole is subtracted.  To order $k^2$
one needs only $\ell=0,1,2$, which give explicit expressions \cite{long}
with low-density expansions
\begin{eqnarray}\label{c11di}
\hat C_{11}^{DI}(\b k)
&=&{xx_1^2\rho_2\over 20T^*\rho_1^2}\Bigl\{\tsty 1-{40\over 27}x_2+
{25\over 21}x_2^2+O(x_2^3)\nonumber\\
&-&{\tsty{5\over 36}}{k^2\over\k^2}\Bigl[\tsty 1+{7\over 15}x_2^2+O(x_2^3)
\Bigr]+O(k^4)\Bigl\}, 
\end{eqnarray}
\begin{eqnarray}\label{c12di}
\hat C_{12}^{DI}(\b k)
&=&-{xx_1^2\over 12T^*x_2\rho_1}\Bigl\{\tsty 1-{6\over 5}x_2
+{8\over 9}x_2^2+O(x_2^3)\nonumber\\
&-&\tsty{5\over 18}\ds{k^2\over\k^2}\Bigl[\tsty x_2-{3\over 5}x_2^2+
{1\over 5}x_2^3+O(x_2^4)\Bigr]+O(k^4)\Bigr\},
\end{eqnarray}
where $x_1\equiv\k a_1$ while $\hat C_{22}^{DI}(\b k)=0$.

Finally,  hard-core exclusion may be approximated by local,
free-volume terms $\cF^{HC}=-\sum_i\rho_i\ln\bigl[1-
\sum_j B_j\rho_j\bigr]$
with (i) $B_1={1\over 2}B_2=4a^3/3\sqrt 3$ to yield bcc close-packing
or (ii) $B_1={1\over 2}B_2=2\pi a^3/3$ for the exact ion-ion second
virial coefficient \cite{FL}; or (iii), perhaps preferably, by the local
Carnahan-Starling mixture form \cite{CSmix}
\begin{equation}\label{hc2}
{6\over\pi}\biggl[\Bigl(\zeta_0-{\zeta_2^3\over 
\zeta_3^2}\Bigr)\ln(1-\zeta_3)-{3\zeta_1\zeta_2\over 1-\zeta_3}
-{\zeta_2^3\over\zeta_3(1-\zeta_3)^2}\biggr],
\end{equation}
where $\zeta_n\equiv{1\over 6}\pi\sum_i\rho_i(\b r)\sigma_i^n$ with 
$\sigma_i$ the hard-core diameter of species $i$; we take $\sigma_1^3=
{1\over 2}\sigma_2^3=a^3$.  For densities near critical, only the 
second virial coefficients prove significant.  Being local, the approximations
(i)-(iii) give $\hat C_{ij}^{HC}(\b k)$ independent of $\b k$.  Nonlocal
effects are easily included at the
second-virial-coefficient level; but that changes the 
critical amplitude $\xi_0^+$ by less than 1\%.

\begin{figure}
\narrowtext
\epsfxsize=\hsize\epsfbox{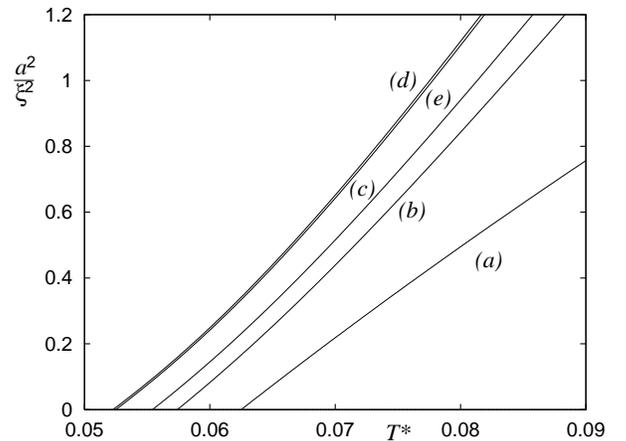}
\caption{Inverse square of the density-density correlation length,
$\xi(T,\rho)$, on the critical isochore according to: (a) pure DH theory;
(b) with Bjerrum association and dipole-pair-ionic-fluid coupling (DHBjDI)
with $a_1=a$, $a_2=1.16198a$; (c)-(e) with hard-core terms
(i)-(iii): see text.}
\end{figure}

For the DHBjDI theories the equilibrium equations require numerical solution.
Fig.\ 1 shows the resulting inverse square correlation lengths {\it vs.\/}
$T^*$ on the critical isochore for various levels of approximation. The 
linear approach of all plots to $\xi^{-2}=0$ represents
the expected classical prediction $\nu={1\over 2}$. The
effects of the DI coupling are less dramatic near $T_c$ than might have
been guessed.
  With the assignments
$a_1=a$, $a_2\simeq 1.162 a$ \cite{FL} the critical
amplitudes are $\xi_0^+/a\simeq 0.7511$, $0.7776$, $0.8186$,
and $0.8147$ for pure DHBjDI theory and with HC treatments (i)-(iii),
respectively.  Increasing $a_1$ to $1.15a$ lowers $\xi_0^+$ by no
more than 8.3\%.  Similarly, taking $a_2$ to be $1.150a$ leads to a 
reduction of less than 1.1\%. (The corresponding changes in $T_c^*$, 
$\rho_c^*$, etc.\ can be found in [3(b)].)

To compare our results for $\xi_0^+$ with experiments on systems
that might plausibly be modeled by the RPM, one needs not only data
for $\xi_0^+$ \cite{expmt} but also some estimate of the
effective hard-core diameter, $a$.  That might be obtained by
matching the $\rho_c^*$ predictions to experiment.  To that end, we 
re-express our results above as $\xi_0^+\rho_c^{1/3}\simeq 0.2275$,
$0.2302$, $0.2375$, and $0.2368$ (in contrast to 
0.1251 for pure DH theory and 0.1828 for the GMSA \cite{Evans}).

Beyond that one must recall that our theory is classical with $\nu=\nu_{MF}
={1\over 2}$ whereas observations indicate the Ising value $\nu_{Is}\simeq
0.63$ or crossover to that for $t\lesssim t_\times$.  However, for 
3\,-D lattice gases, which are described by $\nu_{Is}$ for
all $t\lesssim 1$, the mean-field estimates for $\xi_0^+$, say $\xi_0^{MF}$,
agree with numerical estimates, say $\xi_0^{Is}$, to within 10\% 
\cite{17} (for various lattice structures).  On the other hand,
if crossover is found, the fits for $t<t_\times$ and $t>t_\times$ should 
roughly obey the matching formula $\xi_0^{Is}/\xi_0^{MF}=(t_\times)^{\nu_{Is}
-\nu_{MF}}$.  Data for Na$+$ND$_3$ with $t_\times
\simeq(7$-$9)\times 10^{-3}$ have been fitted in both regions [2(a)(c)]
and confirm the relation.  For this system we find $\xi_0^{Is}\rho_c^{1/3}
= 0.34\pm 0.03$, which is some 40 to 50\% above our estimates.  For
Pitzer's salt [2(b)(e)] we may postulate a crossover
at $t_\times\simeq 1\times 10^{-4}$ \cite{revs}: this yields 
$\xi_0^{Is}\rho_c^{1/3}=0.21\pm 0.04$ which encompasses our values.  
Tetra-n-butyl-ammonium
picrate in n-tridecanol [2(d)] displays crossover and we find $\xi_0^{Is}
\rho_c^{1/3}\simeq 0.22$, close to our prediction.  Finally, for the same
salt in other solvents [2(f)] quite similar values of $\xi_0^{Is}$ fit 
the turbidity data.  Overall the agreement is encouraging when using
the Ising-fitted amplitudes.  It is puzzling, but perhaps significant, 
that in the mean-field
region outside $t_\times$ the values of $\xi_0^{MF}\rho_c^{1/3}$ are all
larger (by factors of 2-3) than our classical theory
predicts!

In conclusion, we have shown how to calculate density-density
correlations within DH theory and its extensions \cite{FL}.
At low densities the correlation length, $\xi(T,\rho)$, diverges
in unexpected but universal fashion potentially amenable to
experimental check.  In the critical region comparison with experiments
on electrolytes proves instructive and raises further questions.  More
concretely, the present theory  enables the observed
classical-to-Ising crossover to be addressed via the Ginzburg criterion
\cite{FL,GC}.

A naive extension of the functional approach outlined is not sufficient for 
studying the charge correlations at higher densities since, when $\b k\to 0$,
the associated density perturbations, even when infinitesimal, induce a 
macroscopic charge imbalance.  However, approaches which separate out the 
long-range Coulombic contributions \cite{KM,vBF} should lead to progress.

We are indebted to Professor David Chandler for a stimulating comment that
led to this work, to Dr.\ J. M. H. Levelt Sengers and Dr.\ Simone Wiegand
for informative discussions, and to Professors J.\ D.\ Weeks, R.\ Evans,
and G.\ Stell for
helpful comments on the manuscript.  The NSF has supported our research
through Grant No.\ CHE 93-11729.

\end{multicols}

\begin{references}
\bibitem{dipcom}For simplicity the density of an ion pair is taken 
at its center of mass.  This approximation can be corrected, but should 
not matter for small $|\b k|$.
\end{references}
\end{document}